\documentclass[aps,pra,twocolumn,floatfix,nofootinbib]{revtex4-2}
\usepackage[colorlinks]{hyperref}
\usepackage[utf8]{inputenc}
\usepackage{graphicx}
\usepackage{multirow}
\usepackage{amssymb}
\usepackage{amsmath}
\usepackage{microtype}
\usepackage{braket}
\usepackage{xcolor}

\begin{document}

	\title{Reply to the comment on ''Quantum sensor networks as exotic field telescopes for multi-messenger astronomy''}

	\author{Andrei Derevianko}
	\affiliation{Department of Physics, University of Nevada, Reno, Nevada 89557, USA}
	\author{Derek Jackson Kimball}
	\affiliation{Physics Department, California State University - East Bay, Hayward, CA 94542}
	\author{Conner Dailey}
	\affiliation{ Department of Physics and Astronomy, University of Waterloo, West Waterloo, Ontario, Canada }
	
\begin{abstract}
The comment by Stadnik  [arXiv:2111.14351v1] claims that ``back-action'', i.e. interaction of exotic
low-mass fields (ELF) with ordinary matter, ``prevents the multi-messenger
astronomy  on human timescales.''  We strongly disagree
with this blanket claim. This is {\em not a general conclusion},  as Stadnik's statement entirely relies on a specific sign of
the ELF-matter interaction. As we demonstrate, there are coupling constant  ranges when, in fact, the screening effects are irrelevant. In addition, the delay between the  arrival of the ELF and gravitational wave bursts is reduced by the ELF-ordinary matter interaction, improving the discovery reach of our proposed novel, exotic physics, modality in multi-messenger
astronomy.
\end{abstract}
	
\maketitle

The comment by Stadnik~\cite{stadnik2021comment} claims that ``back-action'', i.e. interaction of exotic
low-mass fields (ELF) with ordinary matter, ``prevents the multi-messenger
astronomy  on human timescales.''  We strongly disagree
with this blanket claim. This is {\em not a general conclusion},  as this statement entirely relies on a specific sign of
the ELF-matter interaction. One of us (A.D.) pointed out this counter-argument
to Stadnik in private communications. While we are delighted to see that the
counter-argument was partially incorporated into the submitted comment (albeit without
the proper acknowledgement),  Stadnik continues to insist on his  claims. 
This can be misleading and simply damaging to this nascent
research direction. Below we present our counter-argument which demonstrates that there remains a large parameter space for detecting ELFs.

A generic quadratic ELF-matter interaction portal (c.f. Eqs.(58,59)  of our paper~\cite{dailey2020ELF.Concept}) reads,%
\begin{equation}
\mathcal{L}_{\mathrm{clock}}^{(2)}=-\sum_{X}
\Gamma_{X}^{\left(  2\right)} \, \phi^{2}\mathcal{L}_\mathrm{SM}^{X}, \label{Eq:L2Clock}
\end{equation}
where $\mathcal{L}_\mathrm{SM}^{X}$ are various pieces of the Standard Model
Lagrangian, specifically $\mathcal{L}_\mathrm{SM}^{\gamma}=-F_{\mu\nu}^{2}/4$ and
$\mathcal{L}_\mathrm{SM}^{f}=\sum_{\psi}m_{\psi}\bar{\psi}\psi$. Here and below we use natural units, $\hbar=c=1$.
$\Gamma_{X}^{\left(  2\right)}$ in Eq.~(\ref{Eq:L2Clock}) are
coupling constants and importantly their sign can be both positive and
negative. Stadnik's comment focuses on $\Gamma_{X}^{\left(  2\right)}>0$. Part of the confusion stems from his parameterization
$\Gamma_{X}^{\left(  2\right)  }=+1/\Lambda_{X}^{2}$, where the square of real
$\Lambda_{X}$ obscures the sign of  $\Gamma_{X}^{\left(  2\right)  }$. The
proper relation should have been  $\Gamma
_{X}^{\left(  2\right)  }=\pm1/\Lambda_{X}^{2}$. The choice of sign here is
the key to our counter-argument. Stadnik erroneously
claims that the choice of sign in his Eq.~(1)  is identical to our paper (we
used the parameterization~(\ref{Eq:L2Clock}) with $\Gamma_{X}^{\left(  2\right)}$, see  Eqs.(58,59) of Ref.~\cite{dailey2020ELF.Concept}). 

As explicitly stated in the paper~\cite{dailey2020ELF.Concept}, we ignored effects of Galactic dust on the
propagation and attenuation of the ELF waves. Stadnik focuses on such
``back-action'' effects. When $\mathcal{L}_{\mathrm{clock}}^{(2)}$ is combined
with free-field ELF Lagrangian, one obtains equation of motion $\partial^{\mu
}\partial_{\mu}\phi+\left(  m^{2}-2\sum_{X}\Gamma_{X}^{\left(  2\right)
}\mathcal{L}_\mathrm{SM}^{X}\right)  \phi=0$. Assuming constant background
$\mathcal{\bar{L}}_\mathrm{SM}^{X}$ of ordinary matter Lagrangian densities $\mathcal{L}_\mathrm{SM}^{X}$, solutions are the conventional plane
(or spherical) waves with dispersion relation $k^{2}=\omega^{2}-m^{2}+2\sum
_{X}\Gamma_{X}^{\left(  2\right)  }\mathcal{\bar{L}}_\mathrm{SM}^{X}$.   Introducing
index of refraction (with $\beta\equiv -2\sum_{X}\Gamma_{X}^{\left(  2\right)
}\mathcal{\bar{L}}_\mathrm{SM}^{X}$),
\begin{equation}
n\left(  \omega\right)  =\frac{k}{\omega}=\sqrt{1-\frac{m^{2}+\beta}%
{\omega^{2}}}, \label{Eq:IndRefraction}
\end{equation}
maps this problem into well-understood wave-propagation in electrodynamics~\cite{JacksonEM}. The combination $m^{2}+\beta$ is  $m_\mathrm{eff}^2$ in the Stadnik's comment. $\beta > 0$ corresponds to $\Gamma_{X}^{\left(  2\right)  } >0$ and $\beta < 0$ --- to $\Gamma_{X}^{\left(  2\right)  } <0$.
 The square of the effective mass can be misleading as $m_\mathrm{eff}^2$ can be negative.

Now we quickly recover Stadnik's results, but we  keep  track of the interaction sign, so that it is clear where his blanket claims fail.

By screening effect, Stadnik means that when $m^{2}+\beta>\omega^{2}$ in Eq.~(\ref{Eq:IndRefraction}),
the index of refraction becomes purely imaginary and the ELF wave is
attenuated by the sensor environment. This is identical to  the screening phenomena in plasma physics~\cite{JacksonEM}. In the ultra-relativistic limit of our
paper ($m\ll\omega$), this translates into $\beta>0$. However, in the opposite regime of  $\beta<0$ (corresponding to $\Gamma
_{X}^{\left(  2\right)  } <0$), the argument of the square root in the index of refraction is positive. Then 
the attenuation never occurs and there is no screening by the sensor physical package and by the atmosphere. In this case, there is no reduction in sensitivity.

Another point raised by Stadnik is the increase in the lag time between the
gravitational wave (GW) and ELF bursts due to propagation through interstellar
gas. Once again this only holds for his particular choice of sign of
$\Gamma_{X}^{\left(  2\right)}$. Indeed, group velocity is given by~\cite{JacksonEM}
$1/\left(  n+\omega dn/d\omega\right)  $,
\begin{equation}
v_\mathrm{g}=\sqrt{1-\frac{m^{2}+\beta}{\omega^{2}}}\,.
\end{equation}
Positive $\beta$ (Stadnik case) translate into smaller group velocities and
longer GW-ELF lag time. However, $\beta<0$  leads to increasing $v_\mathrm{g}$ and
shorter GW-ELF lag time, thus opening up a larger ELF discovery reach. 

Formally, if $m^{2}+\beta<0$, $v_{g}>1$ and it  seems that the ELF burst
would propagate faster than the speed of light (tachyonic solutions). This is, of course, 
not the case, as the underlying approximation in introducing the concept of group velocity
breaks down, see relevant discussion in electrodynamics textbook~\cite{JacksonEM}.

Another important point is that, as noted by Stadnik, there are no back-action effects at leading order for ELF signals searched for by magnetometer networks (such as the Global Network of Optical Magnetometers for Exotic physics searches, GNOME~\cite{afach2018characterization}) because of their derivative, spin-dependent nature.  
Back-action effects due to magnetic shielding have been considered in Ref.~\cite{kimball2016magnetic} and are already accounted for in all GNOME analysis. 
Similarly there no back-action effects for clock couplings that are linear in ELFs.

To summarize, we appreciate Stadnik's analysis of ``back-action'' effects.
However, his blanket claims of reduced sensitivity and that the back-action
``prevents the multi-messenger astronomy on human timescales'' are {\em not} general.
As we demonstrated, there is a large parameter space that is not excluded by his
analysis. 

{\bf Competing interests} 
The authors declare no competing interests.

{\bf Author contributions} 
A.D. wrote the reply. D.F.J.K. and C.D. provided comments on the draft.


\end{document}